\documentclass{Interspeech2024}

\usepackage{xcolor}
\usepackage{arydshln}
\usepackage {booktabs}
\usepackage {subcaption}
\usepackage{musicography}
\definecolor{EMgray}{gray}{0.45}

\interspeechcameraready 

\title{M2D-CLAP: Masked Modeling Duo Meets CLAP for Learning\\General-purpose Audio-Language Representation}

\name{Daisuke Niizumi$^{1\dagger}$, Daiki Takeuchi$^1$, Yasunori Ohishi$^1$, Noboru Harada$^1$, Masahiro Yasuda$^1$,\\ Shunsuke Tsubaki$^2$, and Keisuke Imoto$^2$}

\address{$^1$ NTT Corporation, $^2$ Doshisha University, Japan}
\email{$^\dagger$daisuke.niizumi@ntt.com}

\begin{document}

\maketitle

\begin{abstract}
Contrastive language-audio pre-training (CLAP) enables zero-shot (ZS) inference of audio and exhibits promising performance in several classification tasks.
However, conventional audio representations are still crucial for many tasks where ZS is not applicable (e.g., regression problems).
Here, we explore a new representation, a general-purpose audio-language representation, that performs well in both ZS and transfer learning.
To do so, we propose a new method, M2D-CLAP, which combines self-supervised learning Masked Modeling Duo (M2D) and CLAP.
M2D learns an effective representation to model audio signals, and CLAP aligns the representation with text embedding. As a result, M2D-CLAP learns a versatile representation that allows for both ZS and transfer learning.
Experiments show that M2D-CLAP performs well on linear evaluation, fine-tuning, and ZS classification with a GTZAN state-of-the-art of 75.17\%, thus achieving a general-purpose audio-language representation.
\end{abstract}

\vspace{-0.15cm}
\section{Introduction}
The advent of CLIP \cite{CLIP} has had a significant impact on diverse domains and promoted the introduction of various audio-language models (ALMs) in the audio domain \cite{AudioCLIP,Wav2CLIP,CLAP2022,CLAP2023,LAION-CLAP}.
These ALMs have enabled diverse applications, including zero-shot (ZS) classification and audio-to-text/text-to-audio retrieval.

On the other hand, conventional audio models (AMs) and their audio representations also remain essential for tasks that cannot be solved with language.
For example, it is challenging to represent the continuous values that are the prediction targets of regression problems in language. CLAP training data are unlikely to contain the language representations of sounds that appear in specific domains, such as industry and medicine.
Therefore, the task that ZS classification can solve is limited.

This study explores a general-purpose audio-language representation as a new representation that can serve as both an ALM and a conventional AM.
When used as a conventional AM, the representation can serve as audio features for a wide range of tasks, including regression, and when used as an ALM, it serves for various tasks, such as ZS classification.

To achieve this, we propose M2D-CLAP, which combines Masked Modeling Duo~\cite{niizumi2022M2D} (M2D), a self-supervised learning (SSL) method, with learning by CLAP.
M2D is an SSL-based AM that uses masked prediction to pre-train a general-purpose audio representation useful for diverse tasks in transfer learning.
Together with CLAP, it enables ZS inference by learning representations that align with textual representations.

Experiments show high transfer learning performance as well as competitive performance in ZS classification, demonstrating that M2D-CLAP achieves a general-purpose audio-language representation. We validate our proposal with many SOTA ALMs/AMs in a unified test environment and compare their performance.
Our contributions are i) the introduction of a general-purpose audio-language representation, ii) proposal of M2D-CLAP, and iii) extensive validation of our representation compared to many SOTA models. We also release our code and a new caption dataset for future research\footnote{\scriptsize{\url{https://github.com/nttcslab/m2d/tree/master/clap}}}.

\vspace{-0.15cm}
\section{Related Work}\label{sec:related-work}
General-purpose audio representations, proposed in SSL methods such as COLA \cite{saeed2020cola} and BYOL-A \cite{niizumi2023byol-a}, have shown effectiveness in various environmental sound, speech, and music tasks.
Representations pre-trained by supervised learning methods, such as PANNs \cite{kong2020panns}, AST \cite{gong2021ast}, and HTS-AT \cite{Chen2022HTS-AT}, have also shown general-purpose effectiveness in various tasks.
Masked prediction SSL methods have recently shown remarkable performance: SSAST \cite{gong2022ssast}, MAE-AST \cite{Baade2022MAE-AST}, and MSM-MAE \cite{niizumi2022msm-mae} learn through reconstruction tasks, while M2D learns by predicting the representation of masked parts of the input signal.
Methods based on masked prediction have also shown high performance: BEATs \cite{chen2022beats} predict tokenized labels, ATST \cite{Li2023ATST-TALSP} incorporates data augmentations, and CED \cite{dinkel2023ced} distills pre-trained models.
While these AM methods are effective in transfer learning, they cannot be applied to ZS classification.

Following CLIP \cite{CLIP}, ALM methods capable of ZS audio classification have been actively proposed.
AudioCLIP \cite{AudioCLIP} and Wav2CLIP \cite{Wav2CLIP} learn audio features that align with the trained CLIP multimodal embedding space.
CLAP \cite{CLAP2022,CLAP2023}, LAION-CLAP \cite{LAION-CLAP}, WavCaps \cite{Mei2023WavCaps}, and FLAP \cite{FLAP} take an approach similar to CLIP, wherein a variety of audio-caption pair datasets are used to learn text and audio embedding that align.
LTU \cite{gong2023LTU}, LTU-AS \cite{gong2023LTU-AS}, and Pengi \cite{deshmukh2023Pengi} take a generative approach using large-language models.
These do not have sufficient general-purpose performance, as shown by the experiments in this study.

Approaches similar to the one in this study are SLIP \cite{Mu2022SLIP} in the image domain, which combines SSL and CLIP, and FLAP, which combines MAE \cite{he2022masked} and CLAP. MAE-based SupMAM-CLAP \cite{xin23d_mamclap} distills CLAP. M2D-S \cite{niizumi2023m2d4speech} extends M2D with an extra network for speech.
Unlike the approaches presented above, we learn general-purpose audio-language representations, ready for both transfer and ZS learning.

\vspace{-0.15cm}
\section{Proposed Method}\label{sec:method}
We propose M2D-CLAP that learns general-purpose audio-language representations by combining SSL (M2D) and supervised learning (CLAP).

\subsection{Background: Masked Modeling Duo}\label{sec:m2d}
\begin{figure}[tbp]
  \vspace{-10pt}
  \centering
  \includegraphics[width=1.0\columnwidth]{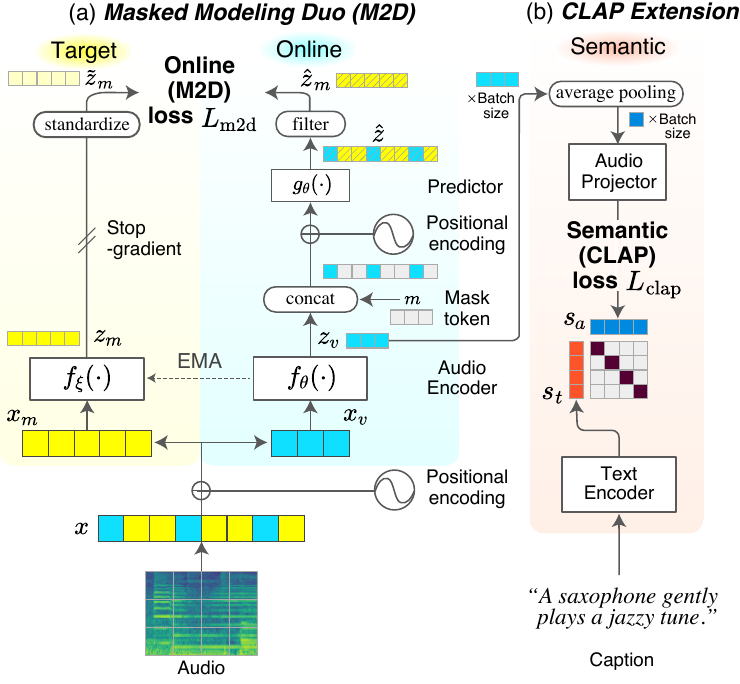}
  \vspace{-15pt}
  \caption{The M2D-CLAP pre-training flow.}
  \label{fig:system}
  \vspace{-10pt}
\end{figure}

M2D is a self-supervised learning framework applicable to 2D structured data input such as images and audio spectrograms, and it trains Vision Transformer \cite{ViT} (ViT) with masked prediction.
As shown in Fig. \ref{fig:system}(a), it consists of two networks, the online and the target, and learns to predict the target output representations using the online output representations.
M2D takes a spectrogram (e.g., 80 frequency bins and 608 time steps) as the input $x$, which is split into patches (e.g., $16\times 16$) and treated as a series (e.g., $(80/16)\times(608/16)=190$ patches).
M2D then adds positional encoding to patches and randomly selects a number of patches according to a masking ratio as masked patches $x_m$ (e.g., 70\% of the input) and the rest as visible patches $x_v$ (e.g., the remaining 30\%).

The online network with a set of weights $\theta$ encodes $x_v$ using the online encoder $f_\theta$ into the representation $z_v = f_\theta(x_v)$.
It concatenates the learnable mask tokens $m$ to $z_v$, adds the position encoding $p$, and inputs them to the predictor $g_\theta$ to predict the representation $\hat{z} = g_\theta(\text{concat}(z_v, m) + p)$.
It then outputs the prediction result $\hat{z}_m = \{\, \hat{z}[i] \mid i \in I_M \,\}$ of the masked patch representations, where $I_M$ is the set of masked patch indices.

The target network defined by parameter $\xi$ outputs the representation $z_m = f_\xi(x_m)$ and standardizes it to the final target output $\tilde{z}_m = ({z_m - \text{mean}{(z_m)}})/{\sqrt{\text{var}{(z_m)}}}$.
The loss is calculated using the online prediction $\hat{z}_m$ against the target output $\tilde{z}_m$ as a training signal by the mean square error (MSE) of $l_2$-normalized $\hat{z}_m$ and $\tilde{z}_m$:
\vspace{-0.1cm}
\begin{equation}
L_\text{m2d} \triangleq ||l_2(\hat{z}_m) - l_2(\tilde{z}_m)||^2_2 = 2 - 2 \cdot \frac{\langle \hat{z}_m, \tilde{z}_m \rangle }{||\hat{z}_m||_2 \cdot ||\tilde{z}_m||_2},
\label{eq:eq-byol-mse}
\vspace{-0.1cm}
\end{equation}
where $\langle\cdot, \cdot\rangle$ denotes the inner product.

The M2D framework updates $\theta$ only to minimize the loss $L_\text{m2d}$ as depicted by the stop-gradient in Fig. \ref{fig:system}(a), and updates $\xi \leftarrow \alpha \xi + (1 - \alpha) \theta$ as an exponential moving average of $\theta$ with a decay rate $\alpha$.
M2D exploits the momentum encoder to learn effective representations from the target network.

\subsection{M2D-CLAP}\label{sec:m2d-clap}
M2D-CLAP performs a multitask learning of M2D and CLAP by adding the CLAP extension shown in Fig. \ref{fig:system}(b) to M2D; M2D-CLAP takes an audio-caption pair as input, feeding audio to M2D and captions to the semantic network in the CLAP extension.
It learns from both the online loss from M2D's masked prediction task and the semantic loss from the CLAP extension.

The semantic network maps audio and captions in a common semantic embedding space.
A text encoder converts captions to a $d_s$ dimensional vector sentence embedding $s_t$, which we use as the semantic embedding as it is.
The audio projector in the network averages the audio visible patch embeddings $z_v$ encoded by M2D and maps them to a $d_s$ dimensional vector semantic embedding $s_a$.

The semantic loss follows CLAP using the cosine similarity $S_{mn}$ between $s_a$ and $s_t$:
\vspace{-0.3cm}
\begin{equation}
S_{mn} = \frac{\langle s_a^{(m)}, s_t^{(n)} \rangle }{||s_a^{(m)}||_2 \cdot ||s_t^{(n)}||_2},
\label{eq:eq-cossim}
\end{equation}
where $s_a^{(m)}$ is the semantic embedding of the $m$th audio batch sample, and $s_t^{(n)}$ is the semantic embedding of the $n$th caption batch sample.
The semantic loss $L_\text{clap}$ is the average of the NT-Xent losses calculated along the audio and caption axes:
\vspace{-0.1cm}
\begin{equation}
L_\text{clap} = \text{\scalebox{0.85}{%
$-\frac{1}{2B} \sum^B_i \left[ \log{\frac{\exp{S_{ii}/\tau}}{\sum^B_j \exp{S_{ji}/\tau}}} + \log{\frac{\exp{S_{ii}/\tau}}{\sum^B_j \exp{S_{ij}/\tau}}} \right],$}}
\label{eq:eq-clip-loss}
\end{equation}
where $B$ is the number of batch samples, and $\tau$ is the learnable temperature parameter. We follow CLIP \cite{CLIP} to initialize $\tau$ with 0.07 and clip it to prevent scaling the logits by more than 100 for training stability.

The entire loss $L$ combines $L_\text{m2d}$ and $L_\text{clap}$:
\vspace{-0.1cm}
\begin{equation}
L = \lambda_\text{m2d} L_\text{m2d} + \lambda_\text{clap} L_\text{clap},
\label{eq:eq-m2d-clap-loss}
\end{equation}
where the loss weights $\lambda_\text{m2d}$ and $\lambda_\text{clap}$ control the contribution.
After pre-training, we transfer only the audio encoder and projector to downstream tasks; the encoder output is for transfer learning, and the projector output is for ZS learning.

Unlike other methods, sentence embedding is used as a multimodal common semantic embedding space to map the audio embedding.
This is beneficial for training an audio embedding to align with the existing semantic embedding space when the space is rich or versatile enough to be compatible with other modalities such as images, for example.
In our experiments, we used General Text Embeddings \cite{li2023GTE} (GTE) with fixed weights as a text encoder and an MLP as an audio projector.

\section{Experiments}
We validate our method in the scenarios of transfer learning by linear evaluation (Section \ref{sec:exp-linear}), fine-tuning (Section \ref{sec:exp-finetuning}), and ZS learning (Section \ref{sec:exp-zero-shot}). 

\subsection{Training Dataset}\label{sec:exp-dataset}
We used AudioSet \cite{gemmeke2017audioset} audio data to train M2D-CLAP, as in M2D.
It consists of 2,005,132 samples (5569 h) of 10-s audio from the balanced and unbalanced train segments.

To form paired audio-caption data with AudioSet, we used the large-scale caption dataset Auto-ACD \cite{sun2023autoacd} as the primary data and our newly created caption dataset, AudioCaps Alternative 4 Captions (ACalt4), to provide variations.
Auto-ACD consists of over 1.9M captions. 
We used the AudioSet subset of Auto-ACD, created label-based captions \textit{"The sound of $\langle$labels$\rangle$"} for the missing captions in Auto-ACD, and made a complete paired dataset of our copy of AudioSet.

ACalt4 is another variation of the AudioCaps \cite{kim2019audiocaps} caption dataset (45K samples) for the audio samples in the subset of the AudioSet. ACalt4 provides four captions for each of the 41,785 samples. To build this dataset, we used the images extracted from the video of the AudioSet sample, as in Auto-ACD. We generated the captions through an automatic pipeline that inputs the image captions generated by BLIP-2 \cite{Li2023BLIP2}, and the AudioSet labels and formats them by leveraging ChatGPT\footnote{\scriptsize{\url{https://openai.com/chatgpt}}}.

\subsection{Experimental Setup}\label{sec:exp-setup}
We used the same M2D configurations as in \cite{niizumi2022M2D}, including the use of ViT Base as the encoder, with an input audio duration of 6 s and a fixed masking ratio of 0.7.
For a sentence encoder, we used GTE-base \cite{li2023GTE} from Hugging Face\footnote{\scriptsize{\url{https://huggingface.co/thenlper/gte-base}}}, with the feature dimension $d_s$ of 768, and fixed the weights. The audio projector is a two-layer MLP with a hidden size of 768.

For the audio data, we randomly cropped 6-s audio from a 10-s sample.
We preprocessed audio samples to a log-scaled mel spectrogram with a sampling frequency of 16,000 Hz, window size of 25 ms, hop size of 10 ms, and mel-spaced frequency bins $F=80$ in the range of 50 to 8000 Hz and standardized them with the statistics 
of AudioSet.

This study differs from \cite{niizumi2022M2D} in that we use the statistics from our pre-training with AudioSet when standardizing the spectrograms for each downstream task. Specifically, we used an average of $-7.1$ and a standard deviation of $4.2$ throughout all downstream task evaluations.
We conducted all evaluations using EVAR\footnote{\scriptsize{\url{https://github.com/nttcslab/eval-audio-repr}}} as a unified evaluation platform.

\begin{table}[tb!]
\vspace{-5pt}
\caption{Fine-tuning settings}
\label{tab:exp-general:ft-parameters}
\vspace{-5pt}
\centering
\resizebox{0.8\columnwidth}{!}{%
\begin{tabular}{llllll}
\toprule
Parameter & AS2M & AS20K & ESC-50 & SPCV2 & VC1 \\
\midrule
Learning rate & 2.0 & 0.5 & 0.5 & 0.5 & 0.0005 \\
Batch size & 64 & 64 & 128 & 128 & 64 \\
Optimizer & LARS & SGD & SGD & SGD & AdamW \\
Mixup ratio & 0.5 & 0.3 & 0.0 & 0.3 & 0.0 \\
Random resize crop (RRC) & - & \checkmark & \checkmark & \checkmark & - \\
SpecAugment$^\sharp$ \cite{specaugment} & 30/192 & 30/192 & 15/48 & 30/48 & 30/48 \\
Training epochs (total) & 70 & 200 & 200 & 200 & 50 \\
Training epochs (warm-up) & 15 & 5 & 5 & 5 & 5 \\
Structured Patchout \cite{Koutini2022passt} ratio & 0.5 & 0.5 & 0.5 & 0.5 & 0.0 \\
\bottomrule
\addlinespace[0.05cm]
\multicolumn{6}{l}{$^{\sharp}$ The frequency/time masking parameters.}\\
\end{tabular}
}
\vspace{-5pt}
\end{table}

\begin{table}[tb!]
\caption{ZS caption conversion rules}
\label{tab:exp:zs-rules}
\vspace{-5pt}
\centering
\resizebox{0.8\columnwidth}{!}{%
\begin{tabular}{ll}
\toprule
Task & Rule \\
\midrule
AS \& FSD & \textit{"$\langle$labels$\rangle$ can be heard"}  \\
ESC50 \& US8K & \textit{"$\langle$label$\rangle$ can be heard"}  \\
CREMA-D & \textit{"(angry person talking $|$ someone talking in disgust}  \\
 & \hspace{0.1cm} \textit{$|$ someone talking with a sense of fear}  \\
 & \hspace{0.1cm} \textit{$|$ someone talking happily and joyfully}  \\
 & \hspace{0.1cm} \textit{$|$ someone talking calmly $|$ someone talking sadly) can be heard"}  \\
GTZAN & \textit{"$\langle$label$\rangle$ music can be heard"}  \\
NSynth & \textit{"the musical instrument sound of $\langle$label$\rangle$ can be heard"}  \\
\bottomrule
\addlinespace[0.05cm]
\end{tabular}
}
\vspace{-10pt}
\end{table}

\vspace{0.1cm}
\noindent\textbf{Pre-training details}\hspace{0.2cm}
We followed M2D for all pre-training settings, including a batch size of 2048 and training epochs of 300.
The loss weights for M2D-CLAP were set to 1.0 for $L_\text{m2d}$ and 0.01 for $L_\text{clap}$.
In cases where ACalt4 had captions for an audio sample, five captions were available, one of which was randomly picked for each training step.
Unlike other ALM methods that initialize an audio encoder with pre-trained weight parameters, we pre-trained M2D from scratch.

\vspace{0.1cm}
\noindent\textbf{Linear evaluation details}\hspace{0.2cm}
All evaluation details and downstream tasks are the same as in \cite{niizumi2022M2D,niizumi2023byol-a}. Tasks include ESC-50 \cite{piczak2015esc50}, UrbanSound8K \cite{salamon2014urbansound} (US8K), Speech Commands V2 \cite{speechcommandsv2} (SPCV2), VoxCeleb1 \cite{voxceleb} (VC1), VoxForge \cite{voxforge} (VF), CREMA-D \cite{cao2014cremad} (CRM-D), GTZAN \cite{gt2013gtzan}, NSynth \cite{nsynth2017}, and Pitch Audio Dataset (Surge synthesizer) \cite{turian2021torchsynth}.
All the tasks are classification problems, and all the results are accuracies.

\vspace{0.1cm}
\noindent\textbf{Fine-tuning details}\hspace{0.2cm}
All downstream tasks are the same as in \cite{niizumi2022M2D}. Tasks include ESC-50, SPCV2, and VC1, plus full AudioSet (AS2M) and the subset AudioSet20K (AS20K).
We extended the fine-tuning settings from \cite{niizumi2022M2D}. In addition to Mixup \cite{niizumi2023byol-a,zhang2018mixup}, RRC \cite{niizumi2023byol-a}, and Structured Patchout \cite{Koutini2022passt}, we used SpecAugment \cite{specaugment} for data augmentation.
The positional encoding was interpolated to adjust it to the duration of the audio sample of the task for AS2M, AS20K, and VC1.
The patch embedding layer weights in ViT were fixed to stabilize the fine-tuning \cite{kumar2022finetune} for ESC-50.
Table \ref{tab:exp-general:ft-parameters} summarizes the settings.

\vspace{0.1cm}
\noindent\textbf{ZS evaluation details}\hspace{0.2cm}
The ZS tasks include AudioSet (AS), ESC-50 (ESC), US8K, CREMA-D (CRD), GTZAN (GTZ), NSynth (NS), and a multi-label classification FSD50K \cite{fonseca2020fsd50k} (FSD).
We conducted the ZS classification in the standard procedure. The model's prediction result was obtained as the label with the closest cosine distance between each test sample and the label's caption, and we obtained the accuracy using these prediction results.
Table \ref{tab:exp:zs-rules} summarizes the conversion rule of the captions from task labels.

\begin{table*}[htb!]
\vspace{-15pt}
\caption{Linear evaluation results (\%) with 95\% CI. We evaluated all models under a unified condition except Pengi.}
\label{tab:results-le}
\centering
\vspace{-5pt}
\resizebox{0.7\textwidth}{!}{%
\begin{tabular}{lllllllllll} \toprule
  &  \multicolumn{2}{c}{Env. sound tasks} & \multicolumn{4}{c}{Speech tasks} & \multicolumn{3}{c}{Music tasks} \\
 \cmidrule(lr){2-3} \cmidrule(lr){4-7} \cmidrule(lr){8-10} 
Model (/masking ratio)  &   ESC-50 &    US8K &    SPCV2 &    VC1 &     VF &    CRM-D &    GTZAN &     NSynth &      Surge & Avg.\\
\midrule

\multicolumn{10}{l}{\textit{(Previous studies: Audio models)}}  \\
\addlinespace[0.05cm]

CED \cite{dinkel2023ced} & 97.3{\fontsize{6pt}{6pt} \selectfont $\pm$0.5} & 87.8{\fontsize{6pt}{6pt} \selectfont $\pm$0.2} & 89.0{\fontsize{6pt}{6pt} \selectfont $\pm$0.3} & 35.2{\fontsize{6pt}{6pt} \selectfont $\pm$0.2} & 94.8{\fontsize{6pt}{6pt} \selectfont $\pm$0.1} & 66.1{\fontsize{6pt}{6pt} \selectfont $\pm$1.3} & 42.3{\fontsize{6pt}{6pt} \selectfont $\pm$15.4} & 75.6{\fontsize{6pt}{6pt} \selectfont $\pm$0.5} & 38.9{\fontsize{6pt}{6pt} \selectfont $\pm$0.6} & 69.7{\fontsize{6pt}{6pt} \selectfont $\pm$2.1} \\
BEATs$_\text{iter3}$ \cite{chen2022beats} & 86.9{\fontsize{6pt}{6pt} \selectfont $\pm$1.4} & 84.8{\fontsize{6pt}{6pt} \selectfont $\pm$0.1} & 89.4{\fontsize{6pt}{6pt} \selectfont $\pm$0.1} & 41.4{\fontsize{6pt}{6pt} \selectfont $\pm$0.7} & 94.1{\fontsize{6pt}{6pt} \selectfont $\pm$0.3} & 64.7{\fontsize{6pt}{6pt} \selectfont $\pm$0.8} & 72.6{\fontsize{6pt}{6pt} \selectfont $\pm$4.3} & 75.9{\fontsize{6pt}{6pt} \selectfont $\pm$0.2} & 39.3{\fontsize{6pt}{6pt} \selectfont $\pm$0.4} & 72.1{\fontsize{6pt}{6pt} \selectfont $\pm$0.9} \\
BEATs$_\text{iter3+}$ \cite{chen2022beats} & 95.5{\fontsize{6pt}{6pt} \selectfont $\pm$0.3} & 87.6{\fontsize{6pt}{6pt} \selectfont $\pm$0.3} & 86.7{\fontsize{6pt}{6pt} \selectfont $\pm$0.1} & 37.0{\fontsize{6pt}{6pt} \selectfont $\pm$0.2} & 92.5{\fontsize{6pt}{6pt} \selectfont $\pm$0.1} & 67.6{\fontsize{6pt}{6pt} \selectfont $\pm$1.5} & 84.6{\fontsize{6pt}{6pt} \selectfont $\pm$0.5} & 73.1{\fontsize{6pt}{6pt} \selectfont $\pm$0.4} & 35.7{\fontsize{6pt}{6pt} \selectfont $\pm$0.3} & 73.4{\fontsize{6pt}{6pt} \selectfont $\pm$0.4} \\
ATST-Clip \cite{Li2023ATST-TALSP} & 94.1{\fontsize{6pt}{6pt} \selectfont $\pm$0.6} & 85.8 $\Lsh$ & 95.1 $\Lsh$ & 72.0 $\Lsh$ & 97.6{\fontsize{6pt}{6pt} \selectfont $\pm$0.0} & 68.8{\fontsize{6pt}{6pt} \selectfont $\pm$1.3} & 78.9{\fontsize{6pt}{6pt} \selectfont $\pm$3.5} & 76.2 $\Lsh$ & 32.8{\fontsize{6pt}{6pt} \selectfont $\pm$0.0} & 77.9{\fontsize{6pt}{6pt} \selectfont $\pm$1.1} \\
ATST-Frame \cite{Li2023ATST-TALSP} & 90.9{\fontsize{6pt}{6pt} \selectfont $\pm$0.6} & 85.8 $\Lsh$ & 94.9 $\Lsh$ &\textbf{77.4} $\Lsh$ &\textbf{98.8{\fontsize{6pt}{6pt} \selectfont $\pm$0.3}}& 72.3{\fontsize{6pt}{6pt} \selectfont $\pm$0.7} & 82.9{\fontsize{6pt}{6pt} \selectfont $\pm$6.0} & 75.9 $\Lsh$ & 40.6{\fontsize{6pt}{6pt} \selectfont $\pm$0.2} & 79.9{\fontsize{6pt}{6pt} \selectfont $\pm$1.6} \\
HTS-AT \cite{Chen2022HTS-AT} & 95.7{\fontsize{6pt}{6pt} \selectfont $\pm$0.7} & 83.8{\fontsize{6pt}{6pt} \selectfont $\pm$0.1} & 82.1{\fontsize{6pt}{6pt} \selectfont $\pm$0.3} & 18.1{\fontsize{6pt}{6pt} \selectfont $\pm$0.4} & 82.3{\fontsize{6pt}{6pt} \selectfont $\pm$0.3} & 56.2{\fontsize{6pt}{6pt} \selectfont $\pm$0.6} &\textbf{85.1{\fontsize{6pt}{6pt} \selectfont $\pm$0.5}}& 73.3{\fontsize{6pt}{6pt} \selectfont $\pm$0.8} & 26.3{\fontsize{6pt}{6pt} \selectfont $\pm$0.5} & 67.0{\fontsize{6pt}{6pt} \selectfont $\pm$0.5} \\
\multicolumn{10}{l}{\textit{(Previous studies: Audio-Language models)}}  \\
\addlinespace[0.05cm]
LAION-CLAP \cite{LAION-CLAP} & 97.3{\fontsize{6pt}{6pt} \selectfont $\pm$0.5} & 86.9{\fontsize{6pt}{6pt} \selectfont $\pm$0.5} & 75.9{\fontsize{6pt}{6pt} \selectfont $\pm$0.5} & 13.4{\fontsize{6pt}{6pt} \selectfont $\pm$0.4} & 80.3{\fontsize{6pt}{6pt} \selectfont $\pm$0.2} & 54.6{\fontsize{6pt}{6pt} \selectfont $\pm$1.0} & 84.3{\fontsize{6pt}{6pt} \selectfont $\pm$2.6} & 72.2{\fontsize{6pt}{6pt} \selectfont $\pm$1.1} & 14.8{\fontsize{6pt}{6pt} \selectfont $\pm$0.5} & 64.4{\fontsize{6pt}{6pt} \selectfont $\pm$0.8} \\
CLAP$_{2022}$ \cite{CLAP2022} & 93.8{\fontsize{6pt}{6pt} \selectfont $\pm$0.1} & 84.2{\fontsize{6pt}{6pt} \selectfont $\pm$0.7} & 59.0{\fontsize{6pt}{6pt} \selectfont $\pm$1.1} & 8.9{\fontsize{6pt}{6pt} \selectfont $\pm$0.6} & 75.8{\fontsize{6pt}{6pt} \selectfont $\pm$1.3} & 54.4{\fontsize{6pt}{6pt} \selectfont $\pm$0.8} & 79.3 $\Lsh$ & 68.2{\fontsize{6pt}{6pt} \selectfont $\pm$0.6} & 8.4{\fontsize{6pt}{6pt} \selectfont $\pm$0.7} & 59.1{\fontsize{6pt}{6pt} \selectfont $\pm$0.7} \\
CLAP$_{2023}$ \cite{CLAP2023} &\textbf{97.7{\fontsize{6pt}{6pt} \selectfont $\pm$0.5}}& 88.4{\fontsize{6pt}{6pt} \selectfont $\pm$0.1} & 86.2{\fontsize{6pt}{6pt} \selectfont $\pm$0.8} & 21.1{\fontsize{6pt}{6pt} \selectfont $\pm$0.3} & 89.6{\fontsize{6pt}{6pt} \selectfont $\pm$0.8} & 62.5{\fontsize{6pt}{6pt} \selectfont $\pm$1.8} & 82.3{\fontsize{6pt}{6pt} \selectfont $\pm$0.5} &\textbf{80.5{\fontsize{6pt}{6pt} \selectfont $\pm$0.1}}& 27.2{\fontsize{6pt}{6pt} \selectfont $\pm$0.5} & 70.6{\fontsize{6pt}{6pt} \selectfont $\pm$0.6} \\
Pengi \cite{deshmukh2023Pengi} & 89.15 $\Lsh$ & - & - & - & - & 50.57 $\Lsh$ & 80.0 $\Lsh$ & - & - & - \\
WavCaps \cite{Mei2023WavCaps} & 97.2{\fontsize{6pt}{6pt} \selectfont $\pm$0.3} & 63.6{\fontsize{6pt}{6pt} \selectfont $\pm$0.6} & 73.3{\fontsize{6pt}{6pt} \selectfont $\pm$1.7} & 16.9{\fontsize{6pt}{6pt} \selectfont $\pm$0.2} & 80.0{\fontsize{6pt}{6pt} \selectfont $\pm$1.0} & 58.6{\fontsize{6pt}{6pt} \selectfont $\pm$0.7} & 80.2{\fontsize{6pt}{6pt} \selectfont $\pm$1.3} & 74.4{\fontsize{6pt}{6pt} \selectfont $\pm$0.9} & 21.1{\fontsize{6pt}{6pt} \selectfont $\pm$0.2} & 62.8{\fontsize{6pt}{6pt} \selectfont $\pm$0.8} \\
\midrule
\multicolumn{6}{l}{\textit{(Baseline: Audio models)}}  \\
\addlinespace[0.05cm]
MSM-MAE/0.75 \cite{niizumi2022msm-mae} $^\dagger$ & 89.2{\fontsize{6pt}{6pt} \selectfont $\pm$0.9} & 87.4{\fontsize{6pt}{6pt} \selectfont $\pm$0.2} & 96.0{\fontsize{6pt}{6pt} \selectfont $\pm$0.1} & 73.6{\fontsize{6pt}{6pt} \selectfont $\pm$0.2} & 97.8{\fontsize{6pt}{6pt} \selectfont $\pm$0.2} & 71.2{\fontsize{6pt}{6pt} \selectfont $\pm$0.4} & 79.2{\fontsize{6pt}{6pt} \selectfont $\pm$0.9} & 74.6{\fontsize{6pt}{6pt} \selectfont $\pm$0.9} &\textbf{43.3{\fontsize{6pt}{6pt} \selectfont $\pm$0.3}}& 79.1{\fontsize{6pt}{6pt} \selectfont $\pm$0.5} \\
M2D/0.6 \cite{niizumi2022M2D} $^\dagger$ & 91.6{\fontsize{6pt}{6pt} \selectfont $\pm$0.5} & 87.2{\fontsize{6pt}{6pt} \selectfont $\pm$0.3} &\textbf{96.2{\fontsize{6pt}{6pt} \selectfont $\pm$0.1}}& 75.0{\fontsize{6pt}{6pt} \selectfont $\pm$0.3} & 98.2{\fontsize{6pt}{6pt} \selectfont $\pm$0.1} & 71.4{\fontsize{6pt}{6pt} \selectfont $\pm$0.9} & 83.4{\fontsize{6pt}{6pt} \selectfont $\pm$3.6} & 76.1{\fontsize{6pt}{6pt} \selectfont $\pm$0.1} & 41.7{\fontsize{6pt}{6pt} \selectfont $\pm$0.2} & 80.1{\fontsize{6pt}{6pt} \selectfont $\pm$0.7} \\
M2D/0.7 \cite{niizumi2022M2D} $^\dagger$ & 91.3{\fontsize{6pt}{6pt} \selectfont $\pm$0.6} & 87.6{\fontsize{6pt}{6pt} \selectfont $\pm$0.2} & 96.0{\fontsize{6pt}{6pt} \selectfont $\pm$0.1} & 73.4{\fontsize{6pt}{6pt} \selectfont $\pm$0.2} & 98.3{\fontsize{6pt}{6pt} \selectfont $\pm$0.0} & 73.0{\fontsize{6pt}{6pt} \selectfont $\pm$0.7} & 84.1{\fontsize{6pt}{6pt} \selectfont $\pm$2.7} & 75.7{\fontsize{6pt}{6pt} \selectfont $\pm$0.1} & 42.1{\fontsize{6pt}{6pt} \selectfont $\pm$0.2} & 80.2{\fontsize{6pt}{6pt} \selectfont $\pm$0.5} \\
\multicolumn{6}{l}{\textit{(Ours: Audio-Language model)}}  \\
\addlinespace[0.05cm]
M2D-CLAP/0.7 & 96.3{\fontsize{6pt}{6pt} \selectfont $\pm$0.3} &\textbf{88.8{\fontsize{6pt}{6pt} \selectfont $\pm$0.6}}& 95.8{\fontsize{6pt}{6pt} \selectfont $\pm$0.3} & 70.3{\fontsize{6pt}{6pt} \selectfont $\pm$0.4} & 98.3{\fontsize{6pt}{6pt} \selectfont $\pm$0.1} &\textbf{73.4{\fontsize{6pt}{6pt} \selectfont $\pm$0.2}}& 84.1{\fontsize{6pt}{6pt} \selectfont $\pm$1.5} & 78.0{\fontsize{6pt}{6pt} \selectfont $\pm$0.5} & 42.4{\fontsize{6pt}{6pt} \selectfont $\pm$0.6} &\textbf{80.8{\fontsize{6pt}{6pt} \selectfont $\pm$0.5}}\\

\bottomrule
\addlinespace[0.08cm]
\multicolumn{8}{l}{$\Lsh$ Results quoted from corresponding papers when they are better than ours or unavailable in our test.}\\
\multicolumn{6}{l}{$^\dagger$ Results obtained with the experimental setup in Section \ref{sec:exp-setup}.}\\
\end{tabular}
}
\vspace{-10pt}
\end{table*}

\subsection{Evaluating Frozen Models (Linear Evaluation)}\label{sec:exp-linear}
We evaluated the SOTA audio and audio-language models and the baseline audio models MSM-MAE and M2D.
Note that the evaluation was conducted under a unified platform with publicly available pre-trained weights, as in  \cite{niizumi2022M2D} and \cite{niizumi2023byol-a}, for fair comparison.
The baselines MSM-MAE and M2D were evaluated under the same conditions described in Section \ref{sec:exp-setup}.

The experimental results in Table \ref{tab:results-le} show that M2D-CLAP performs best on two tasks and has the best average results, demonstrating that it is effective as a general-purpose representation.
Compared to the baselines, the performance is significantly improved for ESC-50 and NSynth, indicating the effect of learning from the caption's supervision.
However, performance deteriorates by about 3pp for VC1 (1251 speaker identification), and we confirmed in preliminary experiments that performance drops further with larger $\lambda_{clap}$, suggesting a trade-off between the CLAP and M2D learnings.
Notably, M2D-CLAP performed well on Surge, an 88 MIDI note classification task similar to a regression problem that is tough for ZS inference to solve.
Overall, this experiment validates that M2D-CLAP retains high general-purpose performance with its frozen representations.

Results also show that the performance of ALM representations varies from task to task and is not generally effective.
While ESC-50, US8K, GTZAN, and Nsynth show high performance, the other five tasks show low performance, especially VC1, which is less than 20\% compared to over 70\% of the top performance.
This may indicate that the coverage of linguistic expressions in the current captions is still limited.
Overall, the ALMs' representations underperform the top results by more than 10pp on average, and they are thus considered to be less versatile as frozen representations.

\subsection{Evaluating Fine-tuning Performance}\label{sec:exp-finetuning}
Table \ref{tab:results-ft} shows the results of fine-tuning. Unlike in the other experiments, we obtained only the results for the baseline and our models due to the difficulty of reproducing the results of other methods in fine-tuning.
M2D-CLAP improved results for AS2M, AS20K, and ESC-50. Meanwhile, it degraded VC1 performance, showing the same trend as in the linear evaluation.
However, the performance of VC1 is similar to that of ATST-Clip, indicating that M2D-CLAP retains its general-purpose performance in the results.
Notably, M2D-CLAP requires only a single pre-training to achieve competitive results with SOTA methods involving multi-iteration/model pre-training.

Among the previous ALMs, CLAP's SPCV2 result of 96.8\% underperforms AMs' 98\%+. However, the performance gap is much smaller than that of linear evaluation, indicating a modest effectiveness of ALMs' representations in fine-tuning. 

\subsection{Evaluating ZS Classification Performance}\label{sec:exp-zero-shot}
Table \ref{tab:results-zs} shows the ZS classification results. M2D-CLAP performs poorly on ESC-50 but well on AudioSet and GTZAN. Particularly, it updates the SOTA result on GTZAN.
Although not in a valid ZS scenario, the best performance on AudioSet is likely because it is the only model trained on AudioSet alone among the ones with AS results.
That also explains the poor performance on  ESC-50; CLAP \cite{CLAP2022} reports that their ESC-50 performance has dropped from 82.6\% to 67.15\% by adding the 1.7M AudioSet samples, aligning with our result of 75.45 \%.
Overall, M2D-CLAP showed competitive ZS performance.

\begin{table}[tb!]
\vspace{-5pt}
\caption{Fine-tuning results with 95\% CI. All results of previous studies are quoted from corresponding papers.}
\label{tab:results-ft}
\centering
\vspace{-5pt}
\resizebox{0.9\columnwidth}{!}{%
\begin{tabular}{llllll} \toprule
  & AS2M & AS20K &     ESC-50 &  SPCV2 &       VC1\\
\vspace{-1pt} Model (/masking ratio)  & mAP & mAP &  acc(\%) &  acc(\%) &    acc(\%)\\
\midrule
\multicolumn{6}{l}{\textit{(Previous studies: Audio models)}}  \\
\addlinespace[0.05cm]
CED \cite{dinkel2023ced} $^{\musDoubleSharp}$ &\textbf{50.0}&\textbf{44.0}& 96.65 & - & - \\
BEATs$_\text{iter3}$ \cite{chen2022beats} & 48.0 & 38.3 & 95.6 & 98.3 & - \\
BEATs$_\text{iter3+}$ \cite{chen2022beats} $^{\sharp}$ & 48.6 & 41.8 &\textbf{98.1}& 98.1 & - \\
SupMAM-CLAP \cite{xin23d_mamclap} $^{\sharp}$ & 48.5 & 38.6 & 97.6 & \textbf{98.7} & - \\
ATST-Clip \cite{Li2023ATST-TALSP} & 45.2 & 37.9 & - & 98.0 & 95.5 \\
ATST-Frame \cite{Li2023ATST-TALSP} & 48.0 & 39.0 & - & 98.1 & {97.3}\\
ATST-C2F \cite{Li2023ATST-TALSP} $^{\sharp}$ & 49.7 & 40.5 & - & 98.4 &\textbf{97.5}\\
HTS-AT \cite{Chen2022HTS-AT} & 47.1 & - & 97.0 & 98.0 & - \\
\multicolumn{6}{l}{\textit{(Previous studies: Audio-Language models)}}  \\
\addlinespace[0.05cm]
AudioCLIP \cite{AudioCLIP} & - & - & 97.15 & - & - \\
Wav2CLIP \cite{Wav2CLIP} & - & - & 85.95 & - & - \\
CLAP$_{2022}$ \cite{CLAP2022} & - & - & 96.7 & 96.8 & - \\
\midrule
\multicolumn{6}{l}{\textit{(Baseline: Audio models)}}  \\
\addlinespace[0.05cm]
MSM-MAE/0.75 \cite{niizumi2022msm-mae} $^\dagger$ & 47.4{\fontsize{6pt}{6pt} \selectfont $\pm$0.1} & 37.9{\fontsize{6pt}{6pt} \selectfont $\pm$0.0} & 95.4{\fontsize{6pt}{6pt} \selectfont $\pm$0.1} & 98.4{\fontsize{6pt}{6pt} \selectfont $\pm$0.0} & 96.6{\fontsize{6pt}{6pt} \selectfont $\pm$0.1} \\
M2D/0.6 \cite{niizumi2022M2D} $^\dagger$ & 47.7{\fontsize{6pt}{6pt} \selectfont $\pm$0.2} & 38.4{\fontsize{6pt}{6pt} \selectfont $\pm$0.1} & 95.6{\fontsize{6pt}{6pt} \selectfont $\pm$0.1} &{98.5{\fontsize{6pt}{6pt} \selectfont $\pm$0.1}}& 96.5{\fontsize{6pt}{6pt} \selectfont $\pm$0.1} \\
M2D/0.7 \cite{niizumi2022M2D} $^\dagger$ & 47.9{\fontsize{6pt}{6pt} \selectfont $\pm$0.0} & 38.6{\fontsize{6pt}{6pt} \selectfont $\pm$0.1} & 96.0{\fontsize{6pt}{6pt} \selectfont $\pm$0.2} & 98.4{\fontsize{6pt}{6pt} \selectfont $\pm$0.1} & 96.3{\fontsize{6pt}{6pt} \selectfont $\pm$0.2} \\
\multicolumn{6}{l}{\textit{(Ours: Audio-Language model)}}  \\
\addlinespace[0.05cm]
M2D-CLAP/0.7 & 48.5{\fontsize{6pt}{6pt} \selectfont $\pm$0.1} & 41.8{\fontsize{6pt}{6pt} \selectfont $\pm$0.2} & 97.4{\fontsize{6pt}{6pt} \selectfont $\pm$0.2} & 98.3{\fontsize{6pt}{6pt} \selectfont $\pm$0.1} & 95.5{\fontsize{6pt}{6pt} \selectfont $\pm$0.2} \\

\bottomrule
\addlinespace[0.08cm]
\multicolumn{6}{l}{$^{\sharp}$ Results using multiple pre-trainings/objectives or $^{\musDoubleSharp}$ large models to distill.}\\
\multicolumn{6}{l}{$^\dagger$ Results obtained with the experimental setup in Section \ref{sec:exp-setup}.}\\
\end{tabular}
}
\vspace{-10pt}
\end{table}

\begin{table}[tb!]
\vspace{-5pt}
\caption{ZS classification results. Underlined results used test task data during training \scriptsize{($\neq$ a ZS scenario)}.}
\label{tab:results-zs}
\centering
\vspace{-5pt}
\resizebox{\columnwidth}{!}{%
\begin{tabular}{llllllll} \toprule
  & AS & FSD & ESC &  US8K & CRD & GTZ &  NS \\
\vspace{-1pt} Model & mAP & mAP & acc(\%) & acc(\%) & acc(\%) & acc(\%) & acc(\%)\\
\midrule

AudioCLIP \cite{AudioCLIP} & - & - & 69.40 $\Lsh$ & 68.78 $\Lsh$ & - & - & - \\
Wav2CLIP \cite{Wav2CLIP} & - & 3.02 $\Lsh$ & 41.4 $\Lsh$ & 40.44 $\Lsh$ & - & - & - \\
WavCaps \cite{Mei2023WavCaps} & \underline{19.60} & \textbf{\underline{52.96}} & 94.8 $\Lsh$ & 81.42 & 19.86 & 45.52 & 27.66 \\
LAION-CLAP \cite{LAION-CLAP} & - & \underline{45.85} & 91.0 $\Lsh$ & 77.0 $\Lsh$ & 23.08 & 47.24 & \textbf{35.28} \\
Proto-LC \cite{Kushwaha23ProtoLC} & - & \underline{52} $\Lsh$ &\textbf{96} $\Lsh$& 73 $\Lsh$ & - & - & - \\
CLAP$_{2022}$ \cite{CLAP2022} & \textbf{5.8} $\Lsh$ & \underline{30.24 $\Lsh$} & 82.6 $\Lsh$ & 75.29 & 22.76 & 28.97 & 21.44 \\
CLAP$_{2023}$ \cite{CLAP2023} &  \underline{10.2 $\Lsh$} & \underline{{48.5} $\Lsh$} & 93.90 $\Lsh$ &\textbf{82.3} $\Lsh$ &\textbf{30.0} $\Lsh$ & 58.4 $\Lsh$ &  \textbf{\underline{58.08}} \\
Pengi \cite{deshmukh2023Pengi} &  \underline{16.35 $\Lsh$} & \underline{46.76 $\Lsh$} & 91.95 $\Lsh$ & 71.85 $\Lsh$ & 18.46 $\Lsh$ & 35.25 $\Lsh$ & \underline{ 50.07 $\Lsh$}\\
LTU\cite{gong2023LTU}/-AS\cite{gong2023LTU-AS} &  \underline{18.7 $\Lsh$} & \underline{46.3} $\Lsh$ & 83.1 $\Lsh$ & - & - & 50.3 $\Lsh$ & - \\
JMLA \cite{du2023JMLA} & - & - & - & - & - & 64.82 $\Lsh$ & - \\
\midrule
\addlinespace[0.1cm]
M2D-CLAP/0.7 & \underline{ \textbf{27.24}} & \textbf{40.82} & 75.45 & 72.40 & 17.73 &\textbf{75.17}& 23.39 \\

\bottomrule
\addlinespace[0.08cm]
\multicolumn{8}{l}{$\Lsh$ Results quoted from each paper when they are better than ours or unavailable in our test.}\\
\end{tabular}
}
\vspace{-10pt}
\end{table}

\section{Conclusion}
This study explored a general-purpose audio-language representation ready for both zero-shot inference and conventional transfer learning.
To this end, we proposed M2D-CLAP, which combines CLAP learning with M2D, an SSL method for learning effective general-purpose representations.
In our experiments, M2D-CLAP showed high performance in linear evaluation, fine-tuning, and zero-shot classification scenarios, and we confirmed that it could learn the desired general-purpose audio-language representation.
In particular, M2D-CLAP further improved the performance of the general-purpose representation compared to M2D and updated GTZAN's SOTA performance in the zero-shot classification.
The general-purpose audio-language representation is effective both as an audio-language model and as a conventional audio representation and is expected to be beneficial for many future application tasks.
Our code and dataset are available online for future studies\footnote{\scriptsize{\url{https://github.com/nttcslab/m2d/tree/master/clap}}}.

\bibliographystyle{IEEEtran}
\bibliography{refs}

\begin{thebibliography}{10}
\providecommand{\url}[1]{#1}
\csname url@samestyle\endcsname
\providecommand{\newblock}{\relax}
\providecommand{\bibinfo}[2]{#2}
\providecommand{\BIBentrySTDinterwordspacing}{\spaceskip=0pt\relax}
\providecommand{\BIBentryALTinterwordstretchfactor}{4}
\providecommand{\BIBentryALTinterwordspacing}{\spaceskip=\fontdimen2\font plus
\BIBentryALTinterwordstretchfactor\fontdimen3\font minus \fontdimen4\font\relax}
\providecommand{\BIBforeignlanguage}[2]{{%
\expandafter\ifx\csname l@#1\endcsname\relax
\typeout{** WARNING: IEEEtran.bst: No hyphenation pattern has been}%
\typeout{** loaded for the language `#1'. Using the pattern for}%
\typeout{** the default language instead.}%
\else
\language=\csname l@#1\endcsname
\fi
#2}}
\providecommand{\BIBdecl}{\relax}
\BIBdecl

\bibitem{CLIP}
A.~Radford, J.~W. Kim, C.~Hallacy, A.~Ramesh, G.~Goh, S.~Agarwal, G.~Sastry, A.~Askell, P.~Mishkin, J.~Clark, G.~Krueger, and I.~Sutskever, ``{Learning Transferable Visual Models From Natural Language Supervision},'' in \emph{ICML}, 2021, pp. 8748--8763.

\bibitem{AudioCLIP}
A.~Guzhov, F.~Raue, J.~Hees, and A.~Dengel, ``{Audioclip: Extending Clip to Image, Text and Audio},'' in \emph{ICASSP}, 2022, pp. 976--980.

\bibitem{Wav2CLIP}
H.-H. Wu, P.~Seetharaman, K.~Kumar, and J.~P. Bello, ``{Wav2CLIP: Learning Robust Audio Representations from Clip},'' in \emph{ICASSP}, 2022, pp. 4563--4567.

\bibitem{CLAP2022}
B.~Elizalde, S.~Deshmukh, M.~Al~Ismail, and H.~Wang, ``{CLAP: Learning Audio Concepts From Natural Language Supervision},'' in \emph{ICASSP}.\hskip 1em plus 0.5em minus 0.4em\relax IEEE, 2023.

\bibitem{CLAP2023}
B.~Elizalde, S.~Deshmukh, and H.~Wang, ``{Natural Language Supervision for General-Purpose Audio Representations},'' \emph{arXiv preprint arXiv:2309.05767}, 2023.

\bibitem{LAION-CLAP}
Y.~Wu, K.~Chen, T.~Zhang, Y.~Hui, T.~Berg-Kirkpatrick, and S.~Dubnov, ``{Large-Scale Contrastive Language-Audio Pretraining with Feature Fusion and Keyword-to-Caption Augmentation},'' in \emph{ICASSP}, 2023.

\bibitem{niizumi2022M2D}
D.~Niizumi, D.~Takeuchi, Y.~Ohishi, N.~Harada, and K.~Kashino, ``{Masked Modeling Duo: Learning Representations by Encouraging Both Networks to Model the Input},'' in \emph{ICASSP}, 2023.

\bibitem{saeed2020cola}
A.~Saeed, D.~Grangier, and N.~Zeghidour, ``Contrastive learning of general-purpose audio representations,'' in \emph{ICASSP}, 2021, pp. 3875--3879.

\bibitem{niizumi2023byol-a}
D.~Niizumi, D.~Takeuchi, Y.~Ohishi, N.~Harada, and K.~Kashino, ``{BYOL for Audio: Exploring Pre-trained General-purpose Audio Representations},'' \emph{IEEE/ACM Trans. Audio, Speech, Language Process.}, vol.~31, p. 137–151, 2023.

\bibitem{kong2020panns}
Q.~Kong, Y.~Cao, T.~Iqbal, Y.~Wang, W.~Wang, and M.~D. Plumbley, ``Panns: Large-scale pretrained audio neural networks for audio pattern recognition,'' \emph{IEEE/ACM Trans. Audio, Speech, Language Process.}, vol.~28, pp. 2880--2894, 2020.

\bibitem{gong2021ast}
Y.~Gong, Y.-A. Chung, and J.~Glass, ``{AST: Audio Spectrogram Transformer},'' in \emph{Interspeech}, 2021, pp. 571--575.

\bibitem{Chen2022HTS-AT}
K.~Chen, X.~Du, B.~Zhu, Z.~Ma, T.~Berg-Kirkpatrick, and S.~Dubnov, ``{HTS-AT}: A hierarchical token-semantic audio transformer for sound classification and detection,'' in \emph{ICASSP}, 2022, pp. 646--650.

\bibitem{gong2022ssast}
Y.~Gong, C.-I. Lai, Y.-A. Chung, and J.~Glass, ``{SSAST: Self-Supervised Audio Spectrogram Transformer},'' in \emph{AAAI}, vol.~36, no.~10, 2022, pp. 10\,699--10\,709.

\bibitem{Baade2022MAE-AST}
A.~Baade, P.~Peng, and D.~Harwath, ``{MAE-AST: Masked Autoencoding Audio Spectrogram Transformer},'' in \emph{Interspeech}, 2022, pp. 2438--2442.

\bibitem{niizumi2022msm-mae}
D.~Niizumi, D.~Takeuchi, Y.~Ohishi, N.~Harada, and K.~Kashino, ``{Masked Spectrogram Modeling using Masked Autoencoders for Learning General-purpose Audio Representation},'' in \emph{HEAR (NeurIPS 2021 Competition)}, vol. 166, 2022, pp. 1--24.

\bibitem{chen2022beats}
S.~Chen, Y.~Wu, C.~Wang, S.~Liu, D.~Tompkins, Z.~Chen, and F.~Wei, ``{BEATs: Audio Pre-Training with Acoustic Tokenizers},'' in \emph{ICML}, 2023.

\bibitem{Li2023ATST-TALSP}
X.~Li, N.~Shao, and X.~Li, ``{Self-Supervised Audio Teacher-Student Transformer for Both Clip-Level and Frame-Level Tasks},'' \emph{IEEE/ACM Trans. Audio, Speech, Language Process.}, vol.~32, pp. 1336--1351, 2024.

\bibitem{dinkel2023ced}
H.~Dinkel, Y.~Wang, Z.~Yan, J.~Zhang, and Y.~Wang, ``{CED: Consistent ensemble distillation for audio tagging},'' in \emph{ICASSP}, 2024.

\bibitem{Mei2023WavCaps}
X.~Mei, C.~Meng, H.~Liu, Q.~Kong, T.~Ko, C.~Zhao, M.~D. Plumbley, Y.~Zou, and W.~Wang, ``{WavCaps: A ChatGPT-Assisted Weakly-Labelled Audio Captioning Dataset for Audio-Language Multimodal Research},'' \emph{arXiv preprint arXiv:2303.17395}, 2023.

\bibitem{FLAP}
C.-F. Yeh, P.-Y. Huang, V.~Sharma, S.-W. Li, and G.~Gosh, ``{FLAP: Fast Language-Audio Pre-Training},'' in \emph{ASRU}, 2023.

\bibitem{gong2023LTU}
Y.~Gong, H.~Luo, A.~H. Liu, L.~Karlinsky, and J.~Glass, ``{Listen, Think, and Understand},'' in \emph{ICLR}, 2024.

\bibitem{gong2023LTU-AS}
Y.~Gong, A.~H. Liu, H.~Luo, L.~Karlinsky, and J.~Glass, ``{Joint Audio and Speech Understanding},'' in \emph{ASRU}, 2023.

\bibitem{deshmukh2023Pengi}
S.~Deshmukh, B.~Elizalde, R.~Singh, and H.~Wang, ``{Pengi: An Audio Language Model for Audio Tasks},'' in \emph{NeurIPS}, 2023.

\bibitem{Mu2022SLIP}
N.~Mu, A.~Kirillov, D.~Wagner, and S.~Xie, ``{SLIP: Self-supervision Meets Language-Image Pre-training},'' in \emph{ECCV}, 2022, pp. 529--544.

\bibitem{he2022masked}
K.~He, X.~Chen, S.~Xie, Y.~Li, P.~Dollár, and R.~Girshick, ``Masked autoencoders are scalable vision learners,'' in \emph{CVPR}, 2022.

\bibitem{xin23d_mamclap}
Y.~Xin, X.~Peng, and Y.~Lu, ``{Masked Audio Modeling with CLAP and Multi-Objective Learning},'' in \emph{Interspeech}, 2023, pp. 2763--2767.

\bibitem{niizumi2023m2d4speech}
D.~Niizumi, D.~Takeuchi, Y.~Ohishi, N.~Harada, and K.~Kashino, ``{Masked Modeling Duo for Speech: Specializing General-Purpose Audio Representation to Speech using Denoising Distillation},'' in \emph{Interspeech}, 2023, pp. 1294--1298.

\bibitem{ViT}
A.~Dosovitskiy, L.~Beyer, A.~Kolesnikov, D.~Weissenborn, X.~Zhai, T.~Unterthiner, M.~Dehghani, M.~Minderer, G.~Heigold, S.~Gelly, J.~Uszkoreit, and N.~Houlsby, ``An image is worth 16x16 words: Transformers for image recognition at scale,'' in \emph{ICLR}, 2021.

\bibitem{li2023GTE}
Z.~Li, X.~Zhang, Y.~Zhang, D.~Long, P.~Xie, and M.~Zhang, ``Towards general text embeddings with multi-stage contrastive learning,'' \emph{arXiv preprint arXiv:2308.03281}, 2023.

\bibitem{gemmeke2017audioset}
J.~F. Gemmeke, D.~P.~W. Ellis, D.~Freedman, A.~Jansen, W.~Lawrence, R.~C. Moore, M.~Plakal, and M.~Ritter, ``{Audio Set}: An ontology and human-labeled dataset for audio events,'' in \emph{ICASSP}, 2017, pp. 776--780.

\bibitem{sun2023autoacd}
L.~Sun, X.~Xu, M.~Wu, and W.~Xie, ``A large-scale dataset for audio-language representation learning,'' \emph{arXiv preprint arXiv:2309.11500}, 2023.

\bibitem{kim2019audiocaps}
C.~D. Kim, B.~Kim, H.~Lee, and G.~Kim, ``{AudioCaps: Generating Captions for Audios in The Wild},'' in \emph{NAACL-HLT}, 2019.

\bibitem{Li2023BLIP2}
J.~Li, D.~Li, S.~Savarese, and S.~C.~H. Hoi, ``{BLIP-2: Bootstrapping Language-Image Pre-training with Frozen Image Encoders and Large Language Models},'' in \emph{ICML}, 2023.

\bibitem{specaugment}
D.~S. Park, W.~Chan, Y.~Zhang, C.-C. Chiu, B.~Zoph, E.~D. Cubuk, and Q.~V. Le, ``{SpecAugment: A Simple Data Augmentation Method for Automatic Speech Recognition},'' in \emph{Interspeech}, 2019, pp. 2613--2617.

\bibitem{Koutini2022passt}
K.~Koutini, J.~Schlüter, H.~Eghbal-zadeh, and G.~Widmer, ``Efficient training of audio transformers with patchout,'' \emph{Interspeech}, pp. 2753--2757, 2022.

\bibitem{piczak2015esc50}
K.~J. Piczak, ``{ESC}: {Dataset} for {Environmental Sound Classification},'' in \emph{ACM-MM}, 2015, pp. 1015--1018.

\bibitem{salamon2014urbansound}
J.~Salamon, C.~Jacoby, and J.~P. Bello, ``A dataset and taxonomy for urban sound research,'' in \emph{ACM-MM}, 2014, pp. 1041--1044.

\bibitem{speechcommandsv2}
P.~{Warden}, ``{Speech Commands: A Dataset for Limited-Vocabulary Speech Recognition},'' \emph{arXiv preprint arXiv::1804.03209}, Apr. 2018.

\bibitem{voxceleb}
A.~Nagrani, J.~S. Chung, and A.~Zisserman, ``Voxceleb: A large-scale speaker identification dataset,'' in \emph{Interspeech}, 2017, pp. 2616--2620.

\bibitem{voxforge}
K.~MacLean, \emph{“Voxforge”}, 2018, available at \url{http://www.voxforge.org/home}.

\bibitem{cao2014cremad}
H.~Cao, D.~G. Cooper, M.~K. Keutmann, R.~C. Gur, A.~Nenkova, and R.~Verma, ``{CREMA-D}: Crowd-sourced emotional multimodal actors dataset,'' \emph{IEEE Trans. Affective Comput.}, vol.~5, no.~4, 2014.

\bibitem{gt2013gtzan}
G.~Tzanetakis and P.~Cook, ``Musical genre classification of audio signals,'' \emph{IEEE Speech Audio Process.}, vol.~10, no.~5, 2002.

\bibitem{nsynth2017}
J.~Engel, C.~Resnick, A.~Roberts, S.~Dieleman, M.~Norouzi, D.~Eck, and K.~Simonyan, ``Neural audio synthesis of musical notes with {W}ave{N}et autoencoders,'' in \emph{ICML}, 2017.

\bibitem{turian2021torchsynth}
J.~Turian, J.~Shier, G.~Tzanetakis, K.~McNally, and M.~Henry, ``One billion audio sounds from {GPU}-enabled modular synthesis,'' in \emph{DAFx2020}, 2021.

\bibitem{zhang2018mixup}
H.~Zhang, M.~Cisse, Y.~N. Dauphin, and D.~Lopez-Paz, ``mixup: Beyond empirical risk minimization,'' in \emph{ICLR}, 2018.

\bibitem{kumar2022finetune}
A.~Kumar, R.~Shen, S.~Bubeck, and S.~Gunasekar, ``{How to Fine-Tune Vision Models with SGD},'' \emph{arXiv preprint arXiv:2211.09359}, 2022.

\bibitem{fonseca2020fsd50k}
E.~Fonseca, X.~Favory, J.~Pons, F.~Font, and X.~Serra, ``{FSD50K: An Open Dataset of Human-Labeled Sound Events},'' \emph{IEEE/ACM Trans. Audio, Speech, Language Process.}, vol.~30, pp. 829--852, 2022.

\bibitem{Kushwaha23ProtoLC}
S.~S. Kushwaha and M.~Fuentes, ``{A multimodal prototypical approach for unsupervised sound classification},'' in \emph{Interspeech}, 2023, pp. 266--270.

\bibitem{du2023JMLA}
X.~Du, Z.~Yu, J.~Lin, B.~Zhu, and Q.~Kong, ``{Joint Music and Language Attention Models for Zero-shot Music Tagging},'' \emph{arXiv preprint arXiv:2310.10159}, 2023.

\end{thebibliography}

\end{document}